\RequirePackage{fix-cm}
\documentclass[twocolumn,epjc3]{svjour3}  
\smartqed  
\RequirePackage{graphicx}
\usepackage[pagewise]{lineno}
\usepackage{booktabs}
\usepackage{amsmath}
\usepackage{amssymb}
\usepackage{multirow}
\usepackage{xcolor}
\usepackage[hidelinks]{hyperref}

\newcommand{\pT}{{p_\text{T}}}
\newcommand{\pTn}[1]{p_\text{T}^{#1}}
\newcommand{\avgpT}[1]{\langle \pTn{#1} \rangle}
\newcommand{\PT}{p'_\text{T}}
\newcommand{\W}{\mathcal{W}}
\newcommand{\Pn}{\mathbb{P}}
\newcommand{\Qn}{\mathcal{R}}
\newcommand{\Pm}{\mathcal{P}_\text{T}}
\newcommand{\coef}{\beta}
\newcommand{\scl}{\lambda}

\journalname{Eur. Phys. J. C}
\begin{document}

\title{
Nonthermal coalescence parameter of light nuclei production in pp collisions at LHC energies\\
}


\author{M. Y. Gallegos Ruiz\thanksref{addr1}
        \and
        D. Rosales Herrera\thanksref{e1,addr1}
        \and
        A. Fern\'andez T\'ellez\thanksref{addr1}
        \and
        J. R. Alvarado Garc\'ia\thanksref{e2,addr1}
        \and
        J. E. Ram\'irez\thanksref{e3,addr1}
}

\thankstext{e1}{e-mail: diana.rosales.herrera@cern.ch}
\thankstext{e2}{e-mail: j.ricardo.alvarado@cern.ch}
\thankstext{e3}{e-mail: jhony.eredi.ramirez.cancino@cern.ch}

\authorrunning{M. Y. Gallegos Ruiz et al.}

\institute{Facultad de Ciencias F\'isico Matem\'aticas, Benem\'erita Universidad Aut\'onoma de Puebla, Apartado Postal 165, 72000 Puebla, Pue., M\'exico \label{addr1}}

\date{Received: date / Accepted: date}


\maketitle

\begin{abstract}
We study the light nucleus production in pp collisions at $\sqrt{s}=7$ and $13$~TeV reported by the ALICE Collaboration within the color string fragmentation framework. 
We adopt the Schwinger mechanism with string tension fluctuations as the particle production mechanism describing the $\pT$ spectrum.
The approximately thermal behavior observed in the light-nucleus spectra motivates the use of a perturbative series expansion of the $q$-Gaussian string tension fluctuations around $q=1$. 
The resulting $\pT$ spectrum is expressed as a perturbative series expansion around the thermal distribution with nonthermal corrections given by Gamma distributions, providing closed forms for the effective temperature, moments, and cumulants. 
In the superstatistical framework, the associated temperature fluctuations show that the medium in which light nuclei are produced is out of equilibrium. 
Additionally, we derive closed formulas for the coalescence parameter at arbitrary order, which agree well with experimental data.
Our results also show that the mass--temperature relation ($T_N/m_N \approx AT_A/m_A$) between the nucleons and nuclei is not preserved in the coalescence process.

\end{abstract}

\section{Introduction}

Light nuclei are composite states formed by a few protons and neutrons, such as deuterons, tritons, and $^{3}\mathrm{He}$.
Their binding energies per nucleon are of one to two orders of magnitude below the temperature of the hadronic medium in which they are produced.
The observed abundances of such states are of several orders of magnitude below nucleon yields, since only a small fraction of emitted nucleons form nuclei.
This low rate of light nuclei production relies on the underlying physics of the production and emission of nucleons. 
Consequently, these abundances constrain the conditions at hadronization and the correlations among the emitted nucleons~\cite{ALICE:2017xrp,STAR:2019sjh,ALICE:2019dgz,ALICE:2019bnp,ALICE:2020foi,ALICE:2021mfm}.
In this way, the study of light-nucleus production provides insights into the late stages of ultrarelativistic collisions.

One way to study the production of light nuclei is to use statistical models, which assume a hadron gas in equilibrium at chemical freeze-out~\cite{Andronic:2010qu,Cleymans:2011pe,Vovchenko:2019kes}.
In this picture, the mass of each species determines its abundance at a given freeze-out temperature and volume.
Then, a common set of freeze-out parameters describes the measured yields of hadrons and light nuclei, with no nucleus-specific parameters.
These models constrain the freeze-out conditions, addressing the integrated yields but not the $\pT$ spectrum.
Alternatively, coalescence models form nuclei from nucleons close in phase space, with a formation probability set by the overlap between the nucleon wave functions and the one of the bound state~\cite{Butler:1963pp,Kapusta:1980zz,Sato:1981ez,Csernai:1986qf,Scheibl:1998tk}. 
In this approach, event generators~\cite{Pierog:2013ria,Sjostrand:2014zea} are combined with a coalescence afterburner that accounts for nucleon correlations and the nuclear wave function to reproduce the $\pT$ spectrum of light nuclei~\cite{Mahlein:2023fmx}. 
In a simple approach, a nucleus with mass number $A$ is formed by $A$ nucleons, each carrying a momentum $\pT=\PT/A$ fixed by momentum conservation.
Therefore, the $\pT$ spectrum of the light nuclei is approximately
\begin{equation}
\W_A(\PT) = B_A(\PT)\left[\W_N \left(\frac{\PT}{A}\right)\right]^{A},
\label{eq:coalescence_intro}
\end{equation}
since the nucleons are emitted independently.
In Eq.~\eqref{eq:coalescence_intro}, $\W_A$ and $\W_N$ denote the $\pT$ spectrum at midrapidity of the light nucleus and of the nucleons, respectively.
Usually, $\W_N$ is the proton $\pT$ spectrum since neutrons and protons belong to the same isospin doublet~\cite{ALICE:2020foi}.
The parameter $B_A$, named the coalescence parameter, measures the probability of forming the nucleus and encodes the source geometry, nucleon correlations, and nuclear structure. 
Under ideal conditions, $B_A$ is inversely proportional to the effective volume of the emission region and remains approximately constant~\cite{ALICE:2019dgz,ALICE:2021mfm}. 

In some cases, the $\pT$ spectrum of nucleons is assumed to follow an exponential distribution, \\$\exp(-\pT/T_N)$, where $T_N$ is identified as the temperature of the medium where they are produced, in analogy with the Boltzmann distribution. 
Following this thermal picture, the $\pT$ spectrum of the nuclei can be written as $[\exp(-\PT/AT_N)]^A$ $=$ $\exp(-\PT/T_N)$. 
Therefore, the nucleus inherits the thermal behavior of the nucleons, implying the mass--temperature relation
\begin{equation}
    A \frac{T_A}{m_A} \approx \frac{T_N}{m_N},
    \label{eq:massT}
\end{equation}
between the nucleus temperature $T_A$ and the nucleon one, with $m_A$ and $m_N$ their respective masses. 

However, the ideal picture described above does not account for a finite size source, collective flow, resonance decays, momentum correlations, or a momentum-dependent probability of forming the nucleus, all of which generate deviations from this baseline~\cite{Scheibl:1998tk,Polleri:1997bp}. 
Indeed, recent deuteron--pion correlation measurements indicate that many deuterons and antideuterons are formed through nuclear reactions following the decay of short-lived baryon resonances~\cite{ALICE:2025byl}. 
Given this, the $\pT$ spectrum of the produced light nuclei may retain an approximately thermal behavior but requires corrections to account for their formation dynamics.

The relation established in Eq.~\eqref{eq:coalescence_intro} requires a framework describing the $\pT$ spectrum of nucleons and light nuclei. 
This observable can be obtained from the color string fragmentation model, which describes particle production from color interactions.
In this model, hadrons are produced by the fragmentation of color strings~\cite{Andersson:1983ia}, with their transverse momentum governed by the Schwinger mechanism~\cite{Schwinger:1951nm}. 
When the string tension is treated as a random variable~\cite{Bialas:1999zg}, Gaussian fluctuations yield a thermal $\pT$ spectrum, which describes the low-$\pT$ region but underestimates the high-$\pT$ production. 
On the other hand, $q$-Gaussian string tension fluctuations introduce the nonextensivity of the system through the parameter $q$, yielding a nonthermal $\pT$ spectrum that reproduces the high-$\pT$ tail~\cite{Pajares:2022uts,Garcia:2023eqg}. 
For light-nucleus production, whose spectra remain close to thermal, $q$ is expected to be close to unity, so the corrections to thermal behavior can be organized as a perturbative series in $\epsilon=q-1$.

In this work, we derive a quasithermal description of the $\pT$ spectrum for systems close to, but not at, the thermal limit, as in the case of the production of light nuclei.
To this end, we derive a perturbative series expansion of the $q$-exponential distribution around $q=1$, which we convolute with the Schwinger mechanism.  
Therefore, the $\pT$ spectrum takes the form of a thermal distribution with nonthermal corrections given by a combination of Gamma distributions.  
From this representation, we obtain closed expressions for the effective temperature, the moments and cumulants of the $\pT$ distribution, the temperature fluctuations, and the coalescence parameter $B_A$.  
We use the quasithermal $\pT$ spectrum to describe the data of proton and light-nucleus production reported by the ALICE Collaboration, highlighting that the medium where light nuclei are produced is out of thermal equilibrium and the nucleon slope is not preserved.

The rest of the paper is organized as follows. 
In Sec.~\ref{sec:quasi}, we derive the perturbative expansion of the $q$-exponential distribution for $0<\epsilon=q-1\ll1$.  
In Sec.~\ref{sec:spectrum}, we obtain the quasithermal $\pT$ spectrum from the convolution of the Schwinger mechanism with the expanded string tension fluctuations, and compute its effective temperature, moments, and cumulants. 
In Sec.~\ref{sec:tempfluc}, we derive the associated temperature fluctuations. 
In Sec.~\ref{sec:coalescence}, we apply the quasithermal $\pT$ spectrum to obtain the coalescence parameter $B_A$. 
In Sec.~\ref{sec:data_analysis}, we present the analysis of the ALICE data.  
Finally, we give our concluding remarks and perspectives in Sec.~\ref{sec:concl}.

\section{Quasithermal distributions}
\label{sec:quasi}

In nature, there are many systems that remain out of thermal equilibrium.
These systems are usually described using Tsallis statistics~\cite{Tsallis:1987eu,Tsallis:2009zex}.
In this framework, the $q$-exponential distribution acquires an important role in describing the probability density function of the random variable of interest.
This distribution is defined as follows
\begin{equation}
e_q(-u) = 
\left[1-(1-q)u\right]^{1/(1-q)},
\label{eq:qexp}
\end{equation}
In Eq.~\eqref{eq:qexp}, $u$ is a variable representing a physical quantity, and $q$ quantifies the nonextensivity degree of the system, measuring how far the system is from equilibrium.
In the limit $q\to1$, Eq.~\eqref{eq:qexp} reduces to the Boltzmann--Gibbs distribution.

Data analyses show that several systems out of equilibrium are characterized by $q$ values very close to 1~\cite{RosalesHerrera:2026pre}, such as the persistence of low-wind periods~\cite{Weber2019}, the interevent times of seismicity catalogs~\cite{Antonopoulos2014}, the delay distributions of railway networks~\cite{Briggs2007}, and the $\pT$ spectra measured in ultrarelativistic collisions~\cite{ALICE:2017xrp,ALICE:2021mfm,ALICE:2010syw,CMS:2010wcx,Cleymans:2011in,Bhattacharyya:2015hya}. 
For these systems, deviations from thermal behavior are small, so it is convenient to expand Eq.~\eqref{eq:qexp} into a perturbative series around $q=1$~\cite{RosalesHerrera:2026pre}.  

\subsection{Generating function and recurrence relations}
\label{sec:genfun}

Let us derive the perturbative expansion around small values of the parameter $\epsilon$ defined as $\epsilon \equiv q - 1$ with $0 < \epsilon \ll 1$.
Note that Eq.~\eqref{eq:qexp} factorizes as
\begin{equation}
e_q(-u) = (1+\epsilon u)^{-1/\epsilon} = \exp(-u) g(u,\epsilon),
\label{eq:qfunctions}
\end{equation}
where $g(u,\epsilon)$ is the generating function
\begin{equation}
g(u,\epsilon) = \exp(u)(1+\epsilon u)^{-1/\epsilon}.
\label{eq:g_def}
\end{equation}
Using the Taylor expansion 
\begin{equation}
\ln(1+z) = \sum_{n=1}^\infty(-1)^{n+1}z^n/n,
\nonumber
\end{equation}
for $|z|<1$, valid for $u < 1/\epsilon$, we obtain
\begin{equation}
\ln g(u,\epsilon) = \sum_{n=2}^\infty (-1)^n \frac{u^n}{n} \epsilon^{n-1},
\end{equation}
which shows that $g(u,\epsilon)$ admits a power series expansion in $\epsilon$ given by
\begin{equation}
g(u,\epsilon) = \sum_{n=0}^\infty \epsilon^n \Pn_n(u),
\label{eq:g_series}
\end{equation}
where $\Pn_n(u)$ are polynomials of degree $2n$ satisfying $\Pn_0(u) = 1$ and $\Pn_n(0) = 0$ for $n \geq 1$. 
The latter condition is derived directly from $g(0,\epsilon) = 1$.

\begin{figure}
    \centering
    \includegraphics{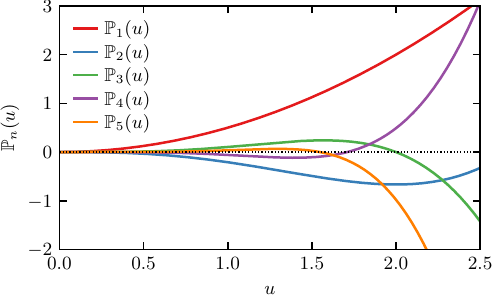}
\caption{Polynomials $\Pn_n(u)$, shown up to $n=5$ as a function of $u$.}
    \label{fig:polinomios_Pn}
\end{figure}

The generating function satisfies the differential equation
\begin{equation}
(1+\epsilon u) \partial_u g = \epsilon u  g.
\label{eq:g_ode}
\end{equation}
Plugging Eq.~\eqref{eq:g_series} into Eq.~\eqref{eq:g_ode} yields the recurrence relation
\begin{equation}
\Pn'_n(u) = u\bigl(\Pn_{n-1}(u) - \Pn'_{n-1}(u)\bigr),
\label{eq:recurrence}
\end{equation}
for $n \geq 1$. 
Primed functions denote derivatives with respect to $u$. 
Integrating Eq.~\eqref{eq:recurrence} with the condition $\Pn_n(0)=0$ gives each polynomial iteratively from the previous one.  
The initial polynomials up to $n=5$ are
\begin{subequations}
\begin{align}
\Pn_0(u) &= 1, \\
\Pn_1(u) &= \frac{u^2}{2}, \\
\Pn_2(u) &= \frac{u^4}{8}-\frac{u^3}{3}, \\
\Pn_3(u) &= \frac{u^6}{48}-\frac{u^5}{6}+\frac{u^4}{4}, \\
\Pn_4(u) &= \frac{u^8}{384}-\frac{u^7}{24}+\frac{13}{72}u^6-\frac{u^5}{5}, \\
\Pn_5(u) &= \frac{u^{10}}{3840}-\frac{u^9}{144}+\frac{17}{288}u^8-\frac{11}{60}u^7+\frac{u^6}{6}.
\end{align}
\label{eq:pol5}
\end{subequations}
In Fig.~\ref{fig:polinomios_Pn}, we show the behavior of the polynomials~\eqref{eq:pol5}.

We can prove by induction that the coefficients of $\Pn_n$, with $n\geq1$, satisfy $\alpha_{n,k}=0$ for $k<n+1$ and $k>2n$, with $k$ the order of the power,\footnote{For $n=1$, $\Pn_1=u^2/2$ contains only $k=2$. If $\Pn_{n-1}$ contains powers $u^k$ with $k\in[n,2(n-1)]$, the right-hand side of Eq.~\eqref{eq:recurrence} raises each power by one, so that after integration the lowest and highest powers of $\Pn_n$ are $u^{n+1}$ and $u^{2n}$, respectively.} so that
\begin{equation}
\Pn_n(u) = \sum_{k=n+1}^{2n}\alpha_{n,k} u^k, \qquad n \geq 1.
\label{eq:pol_structure}
\end{equation}
We substitute Eq.~\eqref{eq:pol_structure} into Eq.~\eqref{eq:recurrence} and compare coefficients of $u^k$, obtaining the recurrence relation for the $\alpha_{n,k}$ coefficients
\begin{equation}
\alpha_{n,k} = \frac{\alpha_{n-1,k-2} - (k-1) \alpha_{n-1,k-1}}{k},
\label{eq:alpha_rec}
\end{equation}
with $\alpha_{0,0} = 1$.

By using the coefficients $\alpha_{n,k}$, the quasithermal approximation of the $q$-exponential distribution~\eqref{eq:qexp}, truncated at an arbitrary order $m$, reads
\begin{equation}
e_q(-u)
\approx
\exp(-u)
\left(
1 + \sum_{n=1}^{m}\epsilon^n
\sum_{k=n+1}^{2n}\alpha_{n,k} u^k
\right).
\label{eq:quasi_thermal}
\end{equation}
Note that the equilibrium term of the expansion~\eqref{eq:quasi_thermal} is the purely exponential distribution.
The terms with $n\geq1$ are the nonthermal corrections, given by functions of the form $\exp(-u)\Pn_n(u)$, each one is a linear combination of Gamma distributions.
These functions satisfy $\int_0^\infty \exp(-u) \Pn_n(u) du=1$ for $n\geq0$.


\section{Quasithermal $\pT$ spectrum}
\label{sec:spectrum}

In the color string fragmentation framework, particle production can be described through the fragmentation of color strings~\cite{Andersson:1983ia}, and the quark--antiquark pair production probability via tunneling is governed by the Schwinger mechanism~\cite{Schwinger:1951nm}. 
For a string of tension $x^2$, this leads to the $\pT$ spectrum of the produced particles
\begin{equation}
\W(\pT) \propto \exp\left(-\frac{\pi \pTn{2}}{x^2}\right).
\label{eq:schwinger}
\end{equation}

To account for stochastic fluctuations in the strength of the color field, $x$ is treated as a random variable with probability density function $P(x)$~\cite{Bialas:1999zg}, and the $\pT$ spectrum is computed as the convolution
\begin{equation}
\W(\pT)
\propto
\int_{0}^{\infty}
\exp\left(-\frac{\pi \pTn{2}}{x^2}\right)
P(x) dx.
\label{eq:convolution}
\end{equation}
If the string tension fluctuates according to a Gaussian distribution, $P_G(x)=\mathcal{N}_G\exp(-x^2/2\sigma^2)$, the resulting $\pT$ spectrum yields a thermal distribution $\W(\pT)\propto\exp(-\pT/\scl)$ with $\scl=\sigma/\sqrt{2\pi}$~\cite{Bialas:1999zg}.
In this case, $\scl$ is identified with the system temperature, by analogy with the Boltzmann distribution~\cite{DiasdeDeus:2006xk}.

However, if the system exhibits small deviations from the Gaussian distribution, the string tension fluctuations can be described by a $q$-Gaussian distribution with
$0<\epsilon = q-1 \ll 1$.
Then, we approximate the $q$-Gaussian distribution to the quasithermal distributions~\eqref{eq:quasi_thermal} with argument $u=x^2/2\sigma^2$, where $x^2$ is the string
tension and $\sigma$ is a parameter that sets the width of the string tension fluctuations.
We substitute this approximation into the convolution~\eqref{eq:convolution}. 
Thus, the $\pT$ spectrum reads
\begin{align}
\W(\pT)
\propto &
\int_0^\infty
\exp\left(-\frac{\pi\pTn{2}}{x^2}\right)
\exp\left(-\frac{x^2}{2\sigma^2}\right)
\nonumber
\\ &
\times
\left(
1+\sum_{n=1}^{m}\epsilon^n\sum_{k=n+1}^{2n}\alpha_{n,k}\frac{x^{2k}}{(2\sigma^2)^{k}}
\right)dx.
\label{eq:convolution_expanded}
\end{align}

To evaluate Eq.~\eqref{eq:convolution_expanded} term by term, we need to compute the integrals
\begin{equation}
I_k
\equiv
\int_0^\infty
\exp \left(-\frac{\pi\pTn{2}}{x^2}\right)
\exp \left(-\frac{x^2}{2\sigma^2}\right)
x^{2k} dx.
\label{eq:Ik_def}
\end{equation}
We use the change of variable $t=x^2$ and the
definition of the modified Bessel function of the second kind~\cite{Gradshteyn2007}
\begin{equation}
\int_0^\infty
t^{\nu-1}
\exp \left(-\beta t - \frac{\gamma}{t}\right)dt
=
2\left(\frac{\gamma}{\beta}\right)^{\nu/2}
K_\nu \left(2\sqrt{\beta\gamma}\right),
\label{eq:Bessel_identity}
\end{equation}
where $\nu=k+1/2$, $\beta=1/2\sigma^2$, and $\gamma=\pi\pTn{2}$. 
This substitution yields
\begin{equation}
I_k
=
(2\pi)^{\frac{k}{2}+\frac{1}{4}}
\sigma^{k+\frac{1}{2}}
\pTn{k+\frac{1}{2}}
K_{k+\frac{1}{2}} \left(\frac{\sqrt{2\pi} \pT}{\sigma}\right).
\label{eq:Ik_K}
\end{equation}
Since $k+1/2$ is a half-integer, $K_{k+1/2}$ becomes~\cite{Gradshteyn2007}
\begin{equation}
K_{k+\frac{1}{2}}(z)
=
\sqrt{\frac{\pi}{2z}} e^{-z}
\sum_{j=0}^k
\frac{(k+j)!}{j! (k-j)!}
\frac{1}{(2z)^j}.
\label{eq:K_half_integer}
\end{equation}
We introduce the scale parameter $\scl=\sigma/\sqrt{2\pi}$, which sets the argument
$z=\pT/\scl$ of the Bessel function.
The integrals~\eqref{eq:Ik_K} become
\begin{equation}
I_k
=
\pi (2\pi)^k \scl^{k+1} 
e^{-\pT/\scl}
\sum_{j=0}^k
\frac{(k+j)!}{j! (k-j)!}
\left(\frac{\scl}{2}\right)^j
\pTn{k-j}.
\label{eq:Ik_explicit}
\end{equation}
For $k=0$ we get $I_0 = \pi \scl e^{-\pT/\scl}$, which has the expected form of the
purely thermal $\pT$ spectrum.
Note that $\scl$ is a scale parameter that does not correspond to the temperature of
the quasithermal $\pT$ spectrum.
We must compute that temperature in the low-$\pT$ limit, as we discuss in
Sec.~\ref{sec:temperature}.

By considering the integrals $I_k$, the $\pT$ spectrum~\eqref{eq:convolution_expanded} reads
\begin{equation}
\W(\pT)
\propto
I_0
+
\sum_{n=1}^{m}
\sum_{k=n+1}^{2n}
\epsilon^n 
\frac{\alpha_{n,k} }{2^k\sigma^{2k}}
I_k,
\label{eq:spectrum_Ik}
\end{equation}
where each $I_k$ carries the same exponential factor $e^{-\pT/\scl}$.
For the sake of notation, we define $y = \pT/\scl$ and the polynomials
\begin{equation}
\Qn_n(y)
=
\sum_{k=n+1}^{2n}
\alpha_{n,k}
\sum_{j=0}^{k}
\frac{(k+j)!}{j! (k-j)!2^{k+j}}
y^{k-j},
\label{eq:Qn_def}
\end{equation}
which are of degree $2n$ in $y$.

By substituting Eq.~\eqref{eq:Ik_explicit} into Eq.~\eqref{eq:spectrum_Ik} and using $2\sigma^2 = 4\pi \scl^2$, we obtain
\begin{equation}
\W(\pT)
\propto
e^{-y}
\left(
1
+
\sum_{n=1}^{m}
\epsilon^n
\Qn_n(y)
\right).
\label{eq:quasithermal_spectrum}
\end{equation}

Note that Eq.~\eqref{eq:quasithermal_spectrum} is a series expansion around the thermal $\pT$ spectrum, where nonthermal perturbative corrections characterize the deviation from equilibrium.
The quasithermal $\pT$ spectrum~\eqref{eq:quasithermal_spectrum} retains the exponential factor $e^{-\pT/\scl}$, modulated by polynomials in $\pT/\scl$ whose coefficients follow from the recurrence relation~\eqref{eq:alpha_rec}. 
Thus, the quasithermal $\pT$ spectrum is a sum of Gamma distributions.

Let us write the first polynomials of the $\pT$ spectrum.
For $n=1$, the only coefficient appearing in Eq.~\eqref{eq:Qn_def} is $\alpha_{1,2}=1/2$, giving
\begin{equation}
\Qn_1(y)
=
\frac{1}{8}
\left(
y^2 + 3y + 3
\right).
\end{equation}
For $n=2$, we use $\alpha_{2,3}=-1/3$ and $\alpha_{2,4}=1/8$, finding
\begin{equation}
\Qn_2(y)
=
\frac{1}{128}
\left(
y^4 + \frac{14}{3}y^3 + 13y^2 + 25y + 25
\right).
\end{equation}
For $n=3$, the coefficients are $\alpha_{3,4}=1/4$, $\alpha_{3,5}=-1/6$, and $\alpha_{3,6}=1/48$, yielding
\begin{align}
\Qn_3(y)
=
\frac{1}{3072}
\big( & 
\nonumber
y^6 + 5y^5 + 18y^4 
\\ 
& + 60y^3 + 165y^2 + 315y + 315
\big).
\end{align}

\begin{figure}
    \centering
    \includegraphics{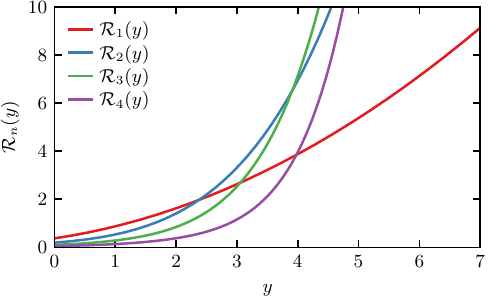}
\caption{Polynomials $\Qn_n(y)$ defined in Eq.~\eqref{eq:Qn_def}, shown up to $n=4$ as a function of the dimensionless variable $y=\pT/\scl$.}
    \label{fig:poli_Qn}
\end{figure}

In Fig.~\ref{fig:poli_Qn}, we plot the polynomials $\Qn_n$ up to $n=4$, which are increasing functions of $y$, meaning the contributions enhance the high-$\pT$ tail of the $\pT$ spectrum relative to the thermal description.

\subsection{Temperature of the quasithermal $\pT$ spectrum}
\label{sec:temperature}

In high-energy physics, it has been observed that the $\pT$ spectrum decays exponentially at low $\pT$.
In this regime, the inverse slope is identified as the temperature of the $\pT$ spectrum, by analogy with the Boltzmann distribution~\cite{DiasdeDeus:2006xk}.
This motivates the following definition of a temperature for a nonthermal distribution
\begin{equation}
T_\text{eff}
=
-\lim_{\pT \to 0} 
\left[\frac{d}{d\pT}\ln \W(\pT) \right]^{-1}.
\label{eq:Teff_def}
\end{equation}
We emphasize that definition~\eqref{eq:Teff_def} applies to $\pT$ spectra that behave as an exponential function at low $\pT$. 
Otherwise, the limit could be ill-defined. 
In the particular case of the thermal distribution, $T_\text{eff}=\scl$, as expected.

We now compute the temperature of the quasithermal $\pT$ spectrum~\eqref{eq:quasithermal_spectrum} using Eq.~\eqref{eq:Teff_def}. 
At $y=0$, the polynomials $\Qn_n(y)$ and their derivatives $\Qn_n'(y)$ satisfy\footnote{From Eq.~\eqref{eq:Qn_def}, the constant term corresponds to $j=k$ and the linear term to $j=k-1$, giving $\Qn_n(0)=\sum_k\alpha_{n,k}(2k)!/(k! 2^{2k})$ and $\Qn_n'(0)=\sum_k\alpha_{n,k}(2k-1)!/((k-1)! 2^{2k-1})$, which are equal since $(2k)!/(k! 2^{2k})=(2k-1)!/((k-1)! 2^{2k-1})$.}
\begin{equation}
\Qn_n(0)
=
\sum_{k=n+1}^{2n}
\alpha_{n,k}
\frac{(2k)!}{k!2^{2k}}
=\Qn_n'(0).
\label{eq:Qn_0}
\end{equation}
This identity simplifies the quotient that results from derivatives in Eq.~\eqref{eq:Teff_def}.
Then, the effective temperature at truncation order $m$ reads
\begin{align}
T_m
&=
\scl
\left(
1
+
\sum_{n=1}^{m}
\epsilon^n
\Qn_n(0)
\right).
\label{eq:Teff}
\end{align}
The nonthermal corrections raise the temperature above the scale parameter $\scl$.  
This increase accompanies the enhancement of the $\pT$ spectrum tail and the flattening of the soft $\pT$ regime as the system departs from thermal equilibrium~\cite{Herrera:2024zjy}, which we corroborate in Sec.~\ref{sec:pTmoments} by computing the statistics of the quasithermal $\pT$ spectrum.



\subsection{Moments and cumulants of the quasithermal $\pT$ spectrum}
\label{sec:pTmoments}

We follow the procedure described in Ref.~\cite{Herrera:2024zjy} for the computation of the moments of the $\pT$ spectrum by introducing
$\left\langle \Pm^{\ell} \right\rangle
= \int_0^\infty \pTn{\ell} \W d\pT / \int_0^\infty \W d\pT$.
Note that the moments reported by the high-energy physics community include the phase-space factor $2\pi\pT$~\cite{Hagedorn:1967tlw}, and are given by
\begin{equation}
\avgpT{\ell}
=
\frac{\left\langle \Pm^{\ell+1}\right\rangle}{\left\langle \Pm \right\rangle}.
\label{eq:moment_def}
\end{equation}

Note that the quasithermal $\pT$ spectrum~\eqref{eq:quasithermal_spectrum} is a linear combination of Gamma distribution terms of the form $y^{\ell}e^{-y}$, which have characteristic function $\varphi(t) = \langle e^{it\pT}\rangle$ of the form  
\begin{equation}
        \varphi_\ell(t) = \left( 1 - i \scl t \right)^{-\ell-1}.
\end{equation}
For the rest of the section, we consider the following parametrization
\begin{equation}
z(t) = 1-it\scl.
\label{eq:zdef}
\end{equation}
Therefore, the characteristic function of the quasithermal $\pT$ spectrum~\eqref{eq:quasithermal_spectrum} is given by
\begin{equation}
\varphi(z)
=
\frac{\mathcal{C}}{z}
\left[
1
+\sum_{n=1}^{m}\epsilon^{n}\sum_{k=n+1}^{2n}\alpha_{n,k}
\sum_{j=0}^{k}\frac{(k+j)!}{j! 2^{k+j} z^{k-j}} 
\right],
\label{eq:charfn}
\end{equation}
with $\mathcal{C}$ being the normalization constant of the quasithermal distribution and $\varphi(z(t=0))=1$.

The moments are obtained from the moment generating function $M=\varphi(z(-it))$.  
Under this substitution, Eq.~\eqref{eq:zdef} yields $z=1-\scl t$, so that $z$ is a real variable, $dz(-it)/dt=-\scl$, and $t=0$ corresponds to $z=1$. 
Then, the moments are obtained as derivatives with respect to $z$ via
\begin{equation}
\left\langle \Pm^{\ell}\right\rangle
=
\left.\frac{d^{\ell}\varphi(z(-it))}{dt^{\ell}}\right|_{t=0}
=
(-\scl)^{\ell}
\left.\frac{d^{\ell}\varphi}{dz^{\ell}}\right|_{z=1}.
\label{eq:moments_z}
\end{equation}

Note that Eq.~\eqref{eq:charfn} is a sum of powers $z^{-(k-j+1)}$, whose derivatives are
$d^{\ell}z^{-(k-j+1)}/dz^{\ell}=(-1)^{\ell}[(k-j+\ell)!/(k-j)!]z^{-(k-j+\ell+1)}$.
Evaluating at $z=1$, we obtain $\left\langle \Pm^{\ell}\right\rangle=\mathcal{C}\scl^{\ell}S_{\ell}$, where
\begin{equation}
S_{\ell}
=
\ell!
+\sum_{n=1}^{m}\epsilon^{n}
\sum_{k=n+1}^{2n}\alpha_{n,k}
\sum_{j=0}^{k}\frac{(k+j)!}{j! (k-j)!}\frac{(\ell+k-j)!}{2^{k+j}}.
\label{eq:Sl}
\end{equation}
In the thermal limit $q\to1$ ($\epsilon\to0$), $\left\langle \Pm^{\ell}\right\rangle$ reduces to
$\left\langle \Pm^{\ell}\right\rangle=\ell!\scl^{\ell}$, recovering the moments of the thermal distribution.

It is possible to probe that the sum over $j$ in Eq.~\eqref{eq:Sl} can be expressed as
\begin{equation}
\sum_{j=0}^{k}\frac{(k+j)!}{j! (k-j)!}\frac{(\ell+k-j)!}{2^{k+j}}
=
\ell! \left(\frac{\ell}{2}+1\right)_{k},
\label{eq:jsum}
\end{equation}
where $(a)_{k}=\Gamma(a+k)/\Gamma(a)$ is the Pochhammer symbol.
Therefore, $S_{\ell}=\ell!\left[1+\sum_{n=1}^{m}\epsilon^{n}\sum_{k=n+1}^{2n}\alpha_{n,k}\left(\frac{\ell}{2}+1\right)_{k}\right]$.

The moments $\avgpT{\ell}$ are obtained by using Eq.~\eqref{eq:moment_def}, which should be truncated at the desired order.
We found
\begin{equation}
\avgpT{\ell}
=
\scl^{\ell}\frac{S_{\ell+1}}{S_{1}}.
\label{eq:pT_moments}
\end{equation}
Expanding the quotient in Eq.~\eqref{eq:pT_moments}, we obtain
\begin{align}
\frac{\avgpT{\ell}}{(\ell+1)! \scl^{\ell}}
=& 1
+ \frac{\ell(\ell+8)}{8} \epsilon
\nonumber \\
&+ \frac{\ell(\ell+2)\left(3\ell^{2}+50\ell+188\right)}{384} \epsilon^{2}
\nonumber \\
&+ \frac{\ell(\ell+2)(\ell+4)(\ell+8)\left(\ell^{2}+18\ell+60\right)}{3072} \epsilon^{3}
\nonumber \\
&+ \mathcal{O}(\epsilon^{4}).
\label{eq:pT_moments_series}
\end{align}

One way to take into account the phase-space factor $2\pi\pT$ in the computation of $ \avgpT{\ell} $ is to consider the probability density function proportional to $\pT\W(\pT)$.
In terms of the characteristic function, multiplying $\W$ by $\pT$ corresponds to differentiating $\varphi$ with respect to $z$, so the generating function associated with $\pT\W$ is proportional to $\varphi'(z)$. 
Since the proportionality factors are independent of $z$, they vanish upon taking derivatives of the natural logarithm.
Using $d/dt=-\scl d/dz$, the cumulants can then be written as
\begin{equation}
\kappa_{\ell}
=
(-\scl)^{\ell}
\left.
\frac{d^{\ell}}{dz^{\ell}}
\ln\left[-\varphi'(z)\right]
\right|_{z=1}.
\label{eq:cumulants_z}
\end{equation}
Thus, the cumulant of order $\ell$ involves derivatives of $\varphi(z)$ up to order $\ell+1$.

Using the derivatives of the logarithm~\cite{comtet1974advanced}, we obtain
\begin{align}
\kappa_{\ell}
= &
\sum_{r=1}^{\ell}(-1)^{r-1}(r-1)!
\nonumber
\\ &
\times
\frac{
B_{\ell,r} \left(
\left\langle\Pm^2\right\rangle,
\left\langle\Pm^3\right\rangle,
\dots,
\left\langle\Pm^{\ell-r+2}\right\rangle
\right)}
{\left\langle\Pm\right\rangle^r},
\label{eq:cumulants_closed}
\end{align}
where $B_{\ell,r}$ are the incomplete Bell polynomials defined as
\begin{equation}
B_{\ell,r}(x_1,\dots,x_{\ell-r+1})
=
\sum_{\{j_i\}}
\frac{\ell!}{j_1! j_2!\cdots}
\prod_{i=1}^{\ell-r+1}
\left(\frac{x_i}{i!}\right)^{j_i},
\label{eq:bell_def}
\end{equation}
and the sum runs over all sets of nonnegative integers $\{j_i\}$ satisfying $\sum_i j_i = r$ and $\sum_i i j_i = \ell$.


Truncating consistently at order $m$, the first cumulants read
\begin{subequations}
\begin{align}
\kappa_1 &= 2\scl\left(1+\tfrac{9}{8}\epsilon+\tfrac{241}{128}\epsilon^2+\cdots\right), \\
\kappa_2 &= 2\scl^2\left(1+3\epsilon+\tfrac{139}{16}\epsilon^2+\cdots\right), \\
\kappa_3 &= 4\scl^3\left(1+\tfrac{45}{8}\epsilon+\tfrac{3177}{128}\epsilon^2+\cdots\right),\\
\kappa_4 &= 12\scl^4\left(1+9\epsilon+56\epsilon^2+\cdots\right),
\end{align}
\label{eq:cumulants}
\end{subequations}
where $\kappa_1=\langle \pT\rangle$ and $\kappa_2=\text{Var}(\pT)$.


The factorized values in Eqs.~\eqref{eq:cumulants} are the cumulants $2(\ell-1)!\scl^{\ell}$ of the distribution $\pT e^{-\pT/\scl}$, which are recovered at $\epsilon \to 0$.
Meanwhile, the deviations from thermal equilibrium enhance $\kappa_\ell$, consistent with the enhanced tails of the $\pT$ spectrum driven by an increased production of high-$\pT$ particles~\cite{Herrera:2024tyq,RosalesHerrera:2026fmg}.

\section{Temperature fluctuations of the quasithermal $\pT$ spectrum}
\label{sec:tempfluc}

In the superstatistics approach, the $\pT$ spectrum is obtained by considering a thermal distribution $e^{-\pT/T}$ with $T$ a random variable. 
Following this approach, we want to express the $\pT$ spectrum as follows~\cite{Wilk:1999dr,Beck:2002lef}
\begin{equation}
\W(\pT)
\propto \int_0^\infty e^{-\pT/T} \mathcal{T}_m(T) dT,
\label{eq:superstatistics}
\end{equation}
where $\mathcal{T}_m(T)$ denotes the distribution of temperature fluctuations, which we derive below for the quasithermal $\pT$ spectrum.  

It is well known that systems at thermal equilibrium have the same temperature throughout. 
In this case, the temperature fluctuations correspond to a Dirac delta.

For the quasithermal $\pT$ spectrum, we start the derivation of the temperature fluctuations by noting that the $n$-th derivative of $e^{-\pT/T}$ with respect to $T$ reads~\cite{Daboul2013}
\begin{equation}
\frac{d^n}{dT^n}e^{-\pT/T}
=
(-1)^n e^{-\pT/T}\sum_{k=1}^{n} L(n,k) (-\pT)^k T^{-n-k},
\label{eq:daboul}
\end{equation}
where $L(n,k)=\binom{n}{k}\binom{n-1}{k-1}(n-k)!$ are the unsigned Lah numbers~\cite{comtet1974advanced,Daboul2013,Riordan1958}.
We invert Eq.~\eqref{eq:daboul} through the orthogonality relation $\sum_k(-1)^{n-k}L(n,k)L(k,m)=\delta_{n,m}$~\cite{Riordan1958} to express
\begin{align}
\left(\frac{\pT}{\scl}\right)^n e^{-\pT/\scl}
=&
\sum_{k=1}^{n}(-1)^{k} L(n,k) \scl^{k}
\nonumber
\\ &
\times
\int_0^\infty \delta^{(k)}(T-\scl) e^{-\pT/T} dT.
\label{eq:master}
\end{align}
Then, we substitute Eq.~\eqref{eq:master} into the quasithermal $\pT$ spectrum, yielding the 
temperature fluctuations $\mathcal{T}_m(T)$ as a superposition of the Dirac delta ($\delta(T-\scl)$) and its $i$-th derivatives ($\delta^{(i)}(T-\scl)$), which reads
\begin{align}
  & \mathcal{T}_m(T) = \delta(T-\scl) 
  + 
  \frac{\scl}{T_m}
\sum_{n=1}^m 
\sum_{k=n+1}^{2n}
\sum_{j=0}^{k-1} 
\nonumber
\\ & \times
\sum_{i=1}^{k-j}
\epsilon^n \alpha_{n,k}
\frac{(k+j)!}{j! (k-j)!2^{k+j}}
(-1)^i L(k-j,i)\scl^i \delta^{(i)}(T-\scl),
\label{eq:Tfluc}  
\end{align}
which is normalized to unity.

The averages over $\mathcal{T}_m(T)$ are computed by using the action of the delta derivatives under integration,
\begin{equation}
\int_0^\infty \psi(T) \delta^{(i)}(T-\scl) dT = (-1)^i \left.\frac{d^i \psi}{dT^i}\right|_{T=\scl},
\label{eq:delta_action}
\end{equation}
with $\psi(T)$ a differentiable function.
Thus, $\langle 1/T \rangle^{-1} = T_m$ and
\begin{subequations}
    \begin{equation}
        \langle T \rangle = \scl \left(
        1
        + \frac{5}{8} \epsilon 
        + \frac{73}{128} \epsilon^2 
        + \frac{575}{1024}  \epsilon^3 
        + \mathcal{O}(\epsilon^4) \right),
    \end{equation}
    \begin{equation}
        \text{Var}(T) = \scl^2 \left(
        \frac{1}{4} \epsilon 
        + \frac{23}{32} \epsilon^2 
        + \frac{197}{128}  \epsilon^3
 + \mathcal{O}(\epsilon^4) \right),
    \end{equation}
    \begin{equation}
        \langle (T - \langle T\rangle)^3 \rangle = \scl^3 \left(
\frac{5}{16} \epsilon ^2
+\frac{223}{128}  \epsilon ^3 
+\frac{12831}{2048}  \epsilon ^4 
        + \mathcal{O}(\epsilon^5) \right).
    \end{equation}    
\end{subequations}

The second and third central moments are nonzero for $m\geq1$ and $m\geq2$, respectively, because higher derivatives of $\delta^{(k)}(T-\scl)$ progressively deform the thermal behavior as more terms of the perturbative expansion are included. 
At first order in $\epsilon$, the temperature fluctuations acquire a finite variance, $\text{Var}(T)=\scl^2\epsilon/4$, while remaining symmetric. 
A nonvanishing variance indicates that the system is not characterized by a single temperature and is out of thermal equilibrium. 
At second order in $\epsilon$, a positive third central moment appears, marking the onset of asymmetric temperature fluctuations quantified by a nonvanishing skewness.

Depending on the system's nonextensivity parameter $\epsilon$, the quasithermal approximation requires more terms in the expansion, which successively incorporate higher derivatives of the Dirac delta function in $\mathcal{T}_m(T)$. 
This description resembles the point-multipole expansion of a charge distribution in classical electrodynamics~\cite{Jackson1999}, where the monopole term plays the role of the equilibrium term and the remaining multipoles introduce one by one the nonthermal corrections in $\mathcal{T}_m(T)$.

\section{Coalescence parameter from the quasithermal $\pT$ spectra}
\label{sec:coalescence}

In the coalescence model, light nuclei are formed at late stages of the collision from nucleons that are close in phase space~\cite{Butler:1963pp,Kapusta:1980zz,Scheibl:1998tk}.
A nucleus forms only when its $A$ nucleons lie close in momentum space, within a range set by a model-dependent parameter.
This parameter is either the coalescence radius, namely the maximum relative momentum for which two nucleons bind~\cite{Butler:1963pp,Kapusta:1980zz}, or the size of the nuclear wave function~\cite{Scheibl:1998tk}.
Momentum conservation fixes the transverse momentum of the nucleus as $\PT=A\pT$, so each nucleon carries the fraction $1/A$ of the nucleus momentum~\cite{ALICE:2017xrp,ALICE:2020foi,ALICE:2021mfm,Butler:1963pp,Scheibl:1998tk}.
Under these conditions, we compute the $\pT$ spectrum of a nucleus with mass number $A$ and electric charge $Z$ as the product of $Z$ proton and $A-Z$ neutron $\pT$ spectra, all evaluated at $\PT/A$.
The coalescence parameter $B_A(\PT)$ sets the proportionality between the nucleus and the nucleon $\pT$ spectra, and quantifies the probability of forming the nucleus.
Additionally, we assume the proton and neutron $\pT$ spectra are equivalent at midrapidity, $\W_p\approx\W_n\approx\W_N$, as commonly adopted in experimental determinations of the coalescence parameters~\cite{ALICE:2017xrp,ALICE:2020foi}.
Several studies report the summed or averaged $\pT$ spectrum of matter and antimatter~\cite{ALICE:2017xrp,ALICE:2015ial,ALICE:2020nkc}, so we identify the proton $\pT$ spectrum with half of the charge-summed one under matter--antimatter symmetry at midrapidity.
We use this identification for the datasets analyzed in Sec.~\ref{sec:data_analysis}.
Then, the $\pT$ spectrum of the light nuclei reads
\begin{equation}
\W_A(\PT) = B_A(\PT)
\left[\W_N \left(\frac{\PT}{A}\right)\right]^{A}.
\label{eq:coalescence_equal}
\end{equation}

For each specie $s$, the quasithermal $\pT$ spectrum truncated at order $m$ is described by Eq.~\eqref{eq:quasithermal_spectrum}, yielding
\begin{equation}
\W_s(\pT)
= \mathcal{A}_s 
e^{-\pT/\scl_s}
\left[1+\sum_{n=1}^{m}
\epsilon_s^{n} \Qn_n \left(\frac{\pT}{\scl_s}\right)\right],
\label{eq:quasithermal_fit}
\end{equation}
where $\mathcal{A}_s$, $\scl_s$, and $\epsilon_s$ are obtained by fitting Eq.~\eqref{eq:quasithermal_fit} to the data.  
We evaluate the nucleus $\pT$ spectrum at $\PT$ and the nucleon spectrum at $\PT/A$, expressed in terms of the scaled variable $y_A=\PT/\scl_A$ for the nucleus and $y_p=\PT/A\scl_p$ for the nucleons. 
Substituting both $\pT$ spectra into the coalescence relation~\eqref{eq:coalescence_equal} and expanding consistently in $\epsilon_A$ and $\epsilon_p$, we obtain
\begin{equation}
B_A(\PT)
=
\frac{\mathcal{A}_A}{\mathcal{A}_p^{A}}
\exp \left(-\frac{\PT}{\scl_0}\right)
\sum_{\substack{i,j\geq0\\i+j\leq m}}
\coef_{i,j} \epsilon_A^{ i} \epsilon_p^{ j},
\label{eq:BA_truncations}
\end{equation}
where we identify the inverse of the constant decay in the coalescence parameter as
$1/\scl_0
=
1/\scl_A - 1/\scl_p$.
The exponential term $\exp(-\PT/\scl_0)$ is proportional to the coalescence parameter for the nucleon and light nuclei in thermal equilibrium.
If coalescence preserves the temperature from nucleons to light nuclei, then $\scl_A=\scl_p$ and $1/\scl_0=0$, making the equilibrium term momentum independent.
Thus, $1/\scl_0$ measures temperature variations during the coalescence process.

The condition $i+j\leq m$ in Eq.~\eqref{eq:BA_truncations} truncates the double expansion at the same order, assuming $\epsilon_A$ and $\epsilon_p$ are of the same order of magnitude. 
The coefficients $\coef_{i,j}$ are listed in Table~\ref{tab:beta}.

\begin{table*}[t]
\centering
\caption{Coefficients $\coef_{i,j}$ of the expansion~\eqref{eq:BA_truncations} up to order $i+j=3$. 
The polynomials $\Qn_n$ are defined in Eq.~\eqref{eq:Qn_def}, with arguments $y_A=\PT/\scl_A$ and $y_p=\PT/A\scl_p$.}
\label{tab:beta}
\begin{tabular}{cccl}
\toprule
$i+j$ & $i$ & $j$ & $\coef_{i,j}$ \\
\midrule
$0$ & 0 & 0 & $1$ \\
\midrule
\multirow{2}{*}{$1$}
 & 1 & 0 & $\Qn_1(y_A)$ \\
 & 0 & 1 & $-A \Qn_1(y_p)$ \\
\midrule
\multirow{3}{*}{$2$}
 & 2 & 0 & $\Qn_2(y_A)$ \\
 & 1 & 1 & $-A \Qn_1(y_A)\Qn_1(y_p)$ \\
 & 0 & 2 & $\dfrac{A(A+1)}{2} \Qn_1^{2}(y_p)-A \Qn_2(y_p)$ \\[6pt]
\midrule
\multirow{4}{*}{$3$}
 & 3 & 0 & $\Qn_3(y_A)$ \\
 & 2 & 1 & $-A \Qn_2(y_A)\Qn_1(y_p)$ \\
 & 1 & 2 & $\Qn_1(y_A)\left[\dfrac{A(A+1)}{2} \Qn_1^{2}(y_p)-A \Qn_2(y_p)\right]$ \\[6pt]
 & 0 & 3 & $-A \Qn_3(y_p)+A(A+1) \Qn_1(y_p)\Qn_2(y_p)-\dfrac{A(A+1)(A+2)}{6} \Qn_1^{3}(y_p)$ \\
\bottomrule
\end{tabular}
\end{table*}

The indices of $\coef_{i,j}$ account for the nonthermal corrections of the $\pT$ spectrum at orders $i$ and $j$ of the nucleus and the nucleons, respectively.
The coefficients $\coef_{i,j}$ are asymmetric because the $\pT$ spectra of nucleons and light nuclei in Eq.~\eqref{eq:coalescence_equal} have different powers.
The nucleus $\pT$ spectrum appears linearly, implying that the pure-nucleus coefficients take the form $\coef_{i,0}=\Qn_i(y_A)$. 
On the other hand, the nucleon $\pT$ spectrum appears to the power $A$ in Eq.~\eqref{eq:coalescence_equal}.
Then, its corrections to the coalescence parameter come from the expansion of $(1+\sum_n\epsilon_p^n\Qn_n)^{-A}$, which produces the combinatorial factors $A$, $A(A+1)/2$, and $A(A+1)(A+2)/6$.

This asymmetry reflects that $A$ nucleons combine into a single nucleus. 
The same asymmetry has a physical counterpart, since $B_A$ is the ratio between the nucleus $\pT$ spectrum and the $A$-th power of the nucleon one. 
When the nucleus $\pT$ spectrum is harder, this ratio grows with $\PT$.
Consequently, forming a nucleus becomes progressively more probable as the transverse momentum increases, in line with the rise of the measured coalescence parameters, usually associated with the decrease of the source homogeneity volume with increasing $\pT$~\cite{Scheibl:1998tk}.

Evaluating the coalescence parameter~\eqref{eq:BA_truncations} for deuterons and $^{3}\mathrm{He}$  yields the corresponding predictions for $B_2$ and $B_3$, respectively.
No additional free parameters are introduced beyond those determined from fits to the $\pT$ spectra.

In the following section, we analyze the $\pT$ spectrum of protons, antideuterons, and $^{3}\mathrm{He}$ reported by the ALICE Collaboration and compare the resulting quasithermal predictions for $B_2$ and $B_3$ with the corresponding experimental measurements.

\section{Data analysis and results}
\label{sec:data_analysis}

\begin{figure}[t]
\centering
\includegraphics{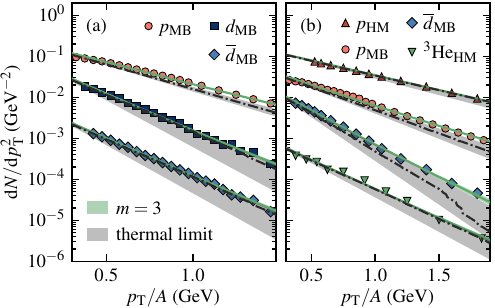}
\caption{Fits of the quasithermal $\pT$ spectrum to the data of proton and light-nucleus production in pp collisions at (a) $\sqrt{s}=7$~TeV and (b) $\sqrt{s}=13$~TeV. 
Markers correspond to the ALICE data~\cite{ALICE:2017xrp,ALICE:2020foi,ALICE:2021mfm,ALICE:2015ial,ALICE:2020nkc}.
The filled region corresponds to the 90\% credible region of the posterior distribution for the quasithermal $\pT$ spectrum at $m=3$ and the thermal limit.
Solid and dash-dotted curves show the most probable value of the quasithermal $\pT$ spectrum and its thermal limit, respectively.
Some data sets were scaled to improve visualization.}
\label{fig:spectra_fits}
\end{figure}

We analyze the $\pT$ spectrum data of protons, (anti)deuterons, and (anti)$^{3}\mathrm{He}$ at $\sqrt{s}=7$ and 13~TeV reported by the ALICE Collaboration~\cite{ALICE:2017xrp,ALICE:2020foi,ALICE:2021mfm,ALICE:2015ial,ALICE:2020nkc}.
To this end, we perform a Bayesian fit of each $\pT$ spectrum with the quasithermal distribution~\eqref{eq:quasithermal_fit} at truncation orders $m=1$, $2$, and $3$ in the $\pT$ window where coalescence is reported, finding the best fits at $m=3$.
In what follows, we report results only for the quasithermal $\pT$ spectrum up to $m=3$.
Table~\ref{tab:CI} summarizes the 90\% credible intervals for the fitted parameters of the quasithermal $\pT$ spectrum. 
In general, we adopted the mode as the measure of central tendency.
However, for the cases of deuterons at 7~TeV, the high-multiplicity protons, and $^{3}\mathrm{He}$, the posterior distribution of $\epsilon$ is a monotonically decreasing function. 
Then, we preferred to use the median for these cases.
Figure~\ref{fig:spectra_fits} shows the 90\% credible interval posterior fits for each dataset, finding good agreement with the experimentally reported data.

\begin{table}[]
    \centering
    \caption{90\% credible intervals for the parameters of the quasithermal distribution at truncation order $m=3$. 
    Data marked with asterisks indicate that the measure of central tendency is the median.
    Otherwise, the measure of central tendency is the mode.
    In parentheses, we denote event classification: (MB) minimum bias and (HM) high multiplicity.}    
\begin{tabular}{lccc}
\hline
Species & $\sqrt{s}$ (TeV) & $\epsilon$ & $\lambda$ (GeV) \\
\hline
$p+\overline{p}$ (MB)  & 7   & $0.115^{+0.039}_{-0.041}$     & $0.37^{+0.02}_{-0.03}$ \\
$d$       (MB)          & 7   & ${0.063^{+0.107}_{-0.057}}^{*}$ & $0.47^{+0.02}_{-0.09}$ \\
$\overline{d}$  (MB)    & 7   & ${0.058^{+0.110}_{-0.053}}^{*}$ & $0.46^{+0.02}_{-0.09}$ \\
$p+\overline{p}$  (MB)   & 13 & $0.116^{+0.057}_{-0.064}$     & $0.39^{+0.04}_{-0.03}$ \\
$\overline{d}$  (MB)      & 13 & $0.090^{+0.091}_{-0.079}$     & $0.47^{+0.05}_{-0.07}$ \\
$p+\overline{p}$ (HM) & 13  & ${0.037^{+0.076}_{-0.034}}^{*}$ & $0.60^{+0.01}_{-0.06}$ \\
${}^{3}\mathrm{He}+{}^{3}\overline{\mathrm{He}}$  (HM) & 13  & ${0.034^{+0.089}_{-0.031}}^{*}$ & $0.90^{+0.05}_{-0.12}$ \\
\hline
\end{tabular}
    \label{tab:CI}
\end{table}


\begin{figure}[t]
\centering
\includegraphics{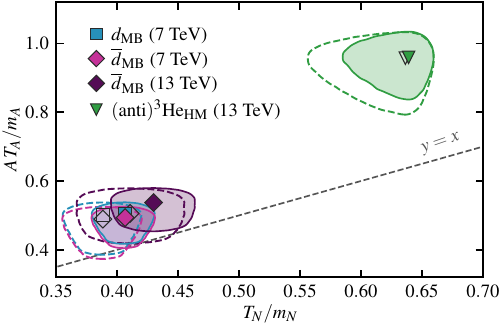}
\caption{$A T_{A}/m_A$ as a function of $T_{N}/m_N$ for different species of light nuclei.
Filled and empty markers correspond to the mode of the joint posterior of the quasithermal descriptions at $m=3$ and their corresponding thermal limits, respectively.
Shaded regions and dashed contours depict the 90\% credible region of the quasithermal description and its thermal limit, respectively.
The dashed line shows the mass--temperature relation $A T_{A}/m_A = T_{N}/m_N$.
}
\label{fig:fit_parameters}
\end{figure}

We test the mass--temperature relation~\eqref{eq:massT} by comparing the scaled temperatures $A T_{A}/m_A$ as a function of $T_p/m_p$, where $m_A$ and $m_p$ denote the nucleus and proton masses, and $T_A$ and $T_p$ are their effective temperature obtained from the fitted parameters via Eq.~\eqref{eq:Teff}.
In Fig.~\ref{fig:fit_parameters}, we show the 90\% credible region of $A T_{A}/m_A$ as a function of $T_{N}/m_N$ for the thermal and quasithermal descriptions at $m=3$.
The four species lie above the line of slope $1$, with posterior probability $P(A T_{A}/m_A > T_{p}/m_p)$ between $0.98$ and $1.00$.
This result indicates a departure from temperature conservation, breaking the mass--temperature relation.
Therefore, the coalescence does not preserve the temperature.

\begin{figure}[t]
\centering
\includegraphics{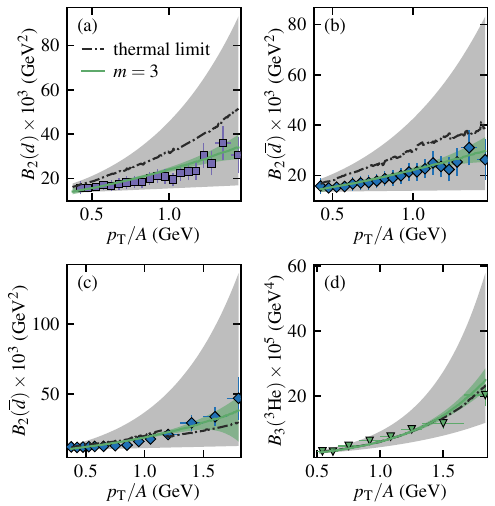}
\caption{Coalescence parameters of the formation of light nuclei in pp collisions: (a) $B_2$ for deuterons at 7~TeV, (b) $B_2$ for antideuterons at 7~TeV, (c) $B_2$ for antideuterons at 13~TeV, and (d) $B_3$ for (anti)$^{3}\mathrm{He}$ at 13~TeV.
Markers represent the ALICE-reported values~\cite{ALICE:2017xrp,ALICE:2020foi,ALICE:2021mfm}.
Shaded regions correspond to the 90\% credible bands, propagated from the independent nucleon and nucleus posteriors.
Solid and dash-dotted curves show the most probable values of the quasithermal description at truncation order $m=3$ and its thermal limit, respectively.
}
\label{fig:coalescence}
\end{figure}

We estimate $B_2$ and $B_3$ by plugging the fitted parameters of the quasithermal $\pT$ spectrum into Eq.~\eqref{eq:BA_truncations}.
The calculations are restricted to the common $\pT$ range reported for the $\pT$ spectra of proton and nucleus production.
Figure~\ref{fig:coalescence} compares the estimated coalescence parameters with ALICE observations, showing a good agreement between the quasithermal estimates and the data reported by the ALICE Collaboration.

\section{Conclusions}
\label{sec:concl}

We derived a perturbative expansion of the $q$-Gaussian string tension fluctuations around the Gaussian limit in the nonextensivity parameter $\epsilon=q-1$. 
Notably, we can obtain the coefficients of the resulting series expansion at arbitrary order via a recurrence relation.
The $\pT$ spectrum resulting from the convolution of the Schwinger mechanism with the series expansion of the $q$-Gaussian string tension fluctuations yields a quasithermal $\pT$ spectrum, given by an equilibrium term and nonthermal corrections that are polynomials in $\pT$, corresponding to a sum of Gamma distributions.
For the quasithermal $\pT$ spectrum, we obtained closed forms for the moments, cumulants, effective temperature, and temperature fluctuations in the superstatistical framework.
The temperature fluctuations have finite variance at first order in $\epsilon$ and nonvanishing skewness at second order.
Therefore, the medium in which light nuclei are produced is out of equilibrium.

We used the quasithermal $\pT$ spectrum to describe the $\pT$ spectra of protons, (anti)deuterons, and (anti)$^{3}\mathrm{He}$ reported by the ALICE Collaboration in pp collisions.
A Bayesian analysis showed that the quasithermal $\pT$ spectrum accurately describes all datasets with the most probable values in the small-$\epsilon$ region.
In all cases, the extracted nonextensivity parameter indicates that light-nucleus production is quasithermal, consistent with the nonextensive characteristics of partonic interactions in charged-particle production~\cite{Pajares:2022uts,Garcia:2023eqg}.
Additionally, the effective temperatures depart from the ideal coalescence relation, $A T_A/m_A\approx T_N/m_N$.
The posterior distributions showed a break in the mass--temperature relation, with significance of $2\sigma$ and $3\sigma$ in the Bayesian analysis for the thermal and quasithermal descriptions, respectively.
Consequently, the coalescence process does not preserve the nucleon temperature in the nucleus $\pT$ spectrum.
This indicates that the coalescence mechanisms redistribute the transverse momentum of the produced nucleons, similar to collective effects producing softened hadrons~\cite{RosalesHerrera:2025zqt}, which may also be considered for improving the quasithermal description.

The results presented in this paper can be extended to analyze the evolution of the coalescence parameter by including information on event shape and other event characteristics, such as multiplicity or centrality.
Furthermore, analyzing heavier nuclei, such as tritons and $^{4}\mathrm{He}$, would test whether the deviation in the mass--temperature relation could be stronger.

%

\begin{acknowledgements}
This work was funded by Secretaría de Ciencia, Humanidades, Tecnología e Innovación (SECIHTI-México) under the project CF-2019/2042, undergraduate fellowship grant number C-2178683/2024, graduate fellowship grant number 1140160, and postdoctoral fellowship grant number 645654.
\end{acknowledgements}

\bibliographystyle{spphys}      
\bibliography{ref}

@article{Andersson:1983ia,
    author = "B. Andersson and G. Gustafson and G. Ingelman and T. Sj{\"o}strand",
    title = "{Parton fragmentation and string dynamics}",
    reportNumber = "LU-TP-83-10",
    doi = "10.1016/0370-1573(83)90080-7",
    journal = "Phys. Rep.",
    volume = "97",
    OPTnumber = "2--3",
    pages = "31--145",
    year = "1983"
}

@article{RosalesHerrera:2026fmg,
    author = "Rosales Herrera, D. and Calder{\'o}n Mu{\~n}oz, J. C. and Alvarado Garc{\'\i}a, J. R. and Fern{\'a}ndez T{\'e}llez, A. and Ram{\'\i}rez, J. E.",
    title = "{Nonextensive Description of Charged-Particle Production in Ultrarelativistic Collisions}",
    doi = "10.3390/e28030298",
    journal = "Entropy",
    volume = "28",
    OPTnumber = "3",
    pages = "298",
    year = "2026"
}

@article{Schwinger:1951nm,
    author = "Schwinger, Julian S.",
    editor = "Milton, K. A.",
    title = "{On Gauge Invariance and Vacuum Polarization}",
    doi = "10.1103/PhysRev.82.664",
    journal = "Phys. Rev.",
    volume = "82",
    OPTnumber = "5",
    pages = "664--679",
    year = "1951"
}

@article{DiasdeDeus:2006xk,
    author = "Dias de Deus, J. and Pajares, C.",
    title = "{Percolation of color sources and critical temperature}",
    doi = "10.1016/j.physletb.2006.10.018",
    journal = "Phys. Lett. B",
    volume = "642",
    OPTnumber = "5--6",
    pages = "455--458",
    year = "2006"
}

@article{Pajares:2022uts,
    author = "Pajares, C. and Ram\'\i{}rez, J. E.",
    title = "{On the relation between the soft and hard parts of the transverse momentum distribution}",
    doi = "10.1140/epja/s10050-023-01170-w",
    journal = "Eur. Phys. J. A",
    volume = "59",
    OPTnumber = "11",
    pages = "250",
    year = "2023"
}

@article{Garcia:2023eqg,
author = "Alvarado Garc\'\i{}a, J. R. and Rosales Herrera, D. and Fierro, P. and Ram\'\i{}rez, J. E. and Fern\'andez T\'ellez, A. and Pajares, C.",
title = {Soft and hard scales of the transverse momentum distribution in the color string percolation model},
doi = {10.1088/1361-6471/acffe1},
publisher = {IOP Publishing},
journal = {J. Phys. G Nucl. Partic.},
    volume = "50",
    OPTnumber = "12",
    pages = "125105",
    year = "2023"
}

@article{Herrera:2024tyq,
    author = "Rosales Herrera, D. and Alvarado Garc\'\i{}a, J. R. and Fern\'andez T\'ellez, A. and Ram\'\i{}rez, J. E. and Pajares, C.",
    title = "{Nonextensivity and temperature fluctuations of the Higgs boson production}",
    doi = "10.1103/PhysRevC.110.015205",
    journal = "Phys. Rev. C",
    volume = "110",
    OPTnumber = "1",
    pages = "015205",
    year = "2024"
}

@article{Bialas:1999zg,
title = {Fluctuations of the string tension and transverse mass distribution},
journal = {Phys. Lett. B},
volume = {466},
OPTnumber = {2--4},
pages = {301--304},
year = {1999},
issn = {0370-2693},
doi = {10.1016/S0370-2693(99)01159-4},
author = {A. Bialas}
}

@article{Hagedorn:1967tlw,
    author = "Hagedorn, R. and Ranft, J.",
    title = "{Statistical thermodynamics of strong interactions at high-energies. 2. Momentum spectra of particles produced in pp-collisions}",
    reportNumber = "CERN-TH-851",
    journal = "Nuovo Cim. Suppl.",
    volume = "6",
    pages = "169--354",
    year = "1968"
}

@article{Herrera:2024zjy,
    author = "Rosales Herrera, D. and Alvarado Garc\'{\i}a, J. R. and Fern\'andez T\'ellez, A. and Ram\'{\i}rez, J. E. and Pajares, C.",
    title = "{Entropy and heat capacity of the transverse momentum distribution for $pp$ collisions at RHIC and LHC energies}",
    doi = "10.1103/PhysRevC.109.034915",
    journal = "Phys. Rev. C",
    volume = "109",
    OPTnumber = "3",
    pages = "034915",
    year = "2024"
}

@article{Beck:2002lef,
    author = "Beck, C. and Cohen, E. G. D.",
    title = "{Superstatistics}",
    doi = "10.1016/S0378-4371(03)00019-0",
    journal = "Physica A",
    volume = "322",
    pages = "267--275",
    year = "2003",
  publisher = {Elsevier}
}

@article{Wilk:1999dr,
    author = "Wilk, G. and W\l{}odarczyk, Z.",
    title = "{Interpretation of the Nonextensivity Parameter $q$ in Some Applications of Tsallis Statistics and L\'evy Distributions}",
    reportNumber = "SINS-PVIII-1999-9",
    doi = "10.1103/PhysRevLett.84.2770",
    journal = "Phys. Rev. Lett.",
    volume = "84",
    OPTnumber = "13",
    pages = "2770--2773",
    year = "2000"
}

@article{Daboul2013,
  author  = {Daboul, Siad and Mangaldan, Jan and Spivey, Michael Z.
             and Taylor, Peter J.},
  title   = "{The Lah Numbers and the $n$th Derivative of $e^{1/x}$}",
  journal = {Math. Mag.},
  volume  = {86},
  OPTnumber = {1},
  pages   = {39--47},
  year    = {2013},
  doi     = {10.4169/math.mag.86.1.039}
}

@book{comtet1974advanced,
  author    = {Comtet, Louis},
  title     = {Advanced Combinatorics: The Art of Finite and Infinite Expansions},
  publisher = {D. Reidel},
  address   = {Dordrecht},
  year      = {1974},
  doi       = {10.1007/978-94-010-2196-8},
  isbn      = {978-90-277-0441-2}
}

@book{Riordan1958,
  author    = {Riordan, John},
  title     = {An Introduction to Combinatorial Analysis},
  publisher = {John Wiley \& Sons},
  address    = {New York},
  year      = {1958}
}

@book{Jackson1999,
  author    = {Jackson, John David},
  title     = {Classical Electrodynamics},
  edition   = {3},
  publisher = {John Wiley \& Sons},
  address   = {New York},
  year      = {1999},
  isbn      = {978-0-471-30932-1}
}

@article{ALICE:2017xrp,
    author = {{ALICE Collaboration}},
    title = "{Production of deuterons, tritons, $^{3}$He nuclei, and their antinuclei in $pp$ collisions at $\mathbf{\sqrt{{\textit s}}}$ = 0.9, 2.76 and 7 TeV}",
    reportNumber = "CERN-EP-2017-255",
    doi = "10.1103/PhysRevC.97.024615",
    journal = "Phys. Rev. C",
    volume = "97",
    OPTnumber = "2",
    pages = "024615",
    year = "2018"
}

@article{ALICE:2021mfm,
    author = {{ALICE Collaboration}},
    title = "{Production of light (anti)nuclei in pp collisions at $\sqrt{s}$ = 13 TeV}",
    reportNumber = "CERN-EP-2021-194",
    doi = "10.1007/JHEP01(2022)106",
    journal = "J. High Energy Phys.",
    volume = "2022",
    pages = "106",
    year = "2022"
}

@article{Scheibl:1998tk,
    author        = {Scheibl, Ruediger and Heinz, Ulrich W.},
    title         = {{Coalescence and flow in ultrarelativistic heavy ion
                     collisions}},
    journal       = {Phys. Rev. C},
    volume        = {59},
    OPTnumber     = {3},
    pages         = {1585--1602},
    year          = {1999},
    doi           = {10.1103/PhysRevC.59.1585}
}

@article{ALICE:2019bnp,
    author = {{ALICE Collaboration}},
    title = "{Multiplicity dependence of light (anti-)nuclei production in p-Pb collisions at $\sqrt{s_{\rm{NN}}}$ = 5.02 TeV}",
    reportNumber = "CERN-EP-2019-120",
    doi = "10.1016/j.physletb.2019.135043",
    journal = "Phys. Lett. B",
    volume = "800",
    pages = "135043",
    year = "2020"
}

@article{Butler:1963pp,
    author = "Butler, S. T. and Pearson, C. A.",
    title = "{Deuterons from High-Energy Proton Bombardment of Matter}",
    doi = "10.1103/PhysRev.129.836",
    journal = "Phys. Rev.",
    volume = "129",
    OPTnumber = "2",
    pages = "836--842",
    year = "1963"
}

@article{Kapusta:1980zz,
    author = "Kapusta, Joseph I.",
    title = "{Mechanisms for deuteron production in relativistic nuclear collisions}",
    doi = "10.1103/PhysRevC.21.1301",
    journal = "Phys. Rev. C",
    volume = "21",
    OPTnumber = "4",
    pages = "1301--1310",
    year = "1980"
}

@article{Sato:1981ez,
    author = "Sato, H. and Yazaki, K.",
    title = "{On the coalescence model for high-energy nuclear reactions}",
    doi = "10.1016/0370-2693(81)90976-X",
    journal = "Phys. Lett. B",
    volume = "98",
    OPTnumber = "3",
    pages = "153--157",
    year = "1981"
}

@article{Csernai:1986qf,
    author = "Csernai, L. P. and Kapusta, Joseph I.",
    title = "{Entropy and cluster production in nuclear collisions}",
    doi = "10.1016/0370-1573(86)90031-1",
    journal = "Phys. Rep.",
    volume = "131",
    OPTnumber = "4",
    pages = "223--318",
    year = "1986"
}

@article{ALICE:2020foi,
    author = {{ALICE Collaboration}},
    title = "{(Anti-)deuteron production in pp collisions at $\sqrt{s}=13 \ \text{TeV}$}",
    reportNumber = "CERN-EP-2020-025",
    doi = "10.1140/epjc/s10052-020-8256-4",
    journal = "Eur. Phys. J. C",
    volume = "80",
    OPTnumber = "9",
    pages = "889",
    year = "2020"
}

@article{ALICE:2019dgz,
    author = {{ALICE Collaboration}},
    title = "{Multiplicity dependence of (anti-)deuteron production in pp collisions at $\sqrt{s}$ = 7 TeV}",
    reportNumber = "CERN-EP-2019-024",
    doi = "10.1016/j.physletb.2019.05.028",
    journal = "Phys. Lett. B",
    volume = "794",
    pages = "50--63",
    year = "2019"
}

@article{STAR:2019sjh,
    author = {{STAR Collaboration}},
    title = "{Beam energy dependence of (anti-)deuteron production in Au + Au collisions at the BNL Relativistic Heavy Ion Collider}",
    doi = "10.1103/PhysRevC.99.064905",
    journal = "Phys. Rev. C",
    volume = "99",
    OPTnumber = "6",
    pages = "064905",
    year = "2019"
}

@article{Polleri:1997bp,
    author = "Polleri, Alberto and Bondorf, Jakob P. and Mishustin, Igor N.",
    title = "{Effects of collective expansion on light cluster spectra in relativistic heavy ion collisions}",
    reportNumber = "NBI-97-44",
    doi = "10.1016/S0370-2693(97)01455-X",
    journal = "Phys. Lett. B",
    volume = "419",
    OPTnumber = "1--4",
    pages = "19--24",
    year = "1998"
}

@article{Mahlein:2023fmx,
    author = "Mahlein, Maximilian and Barioglio, Luca and Bellini, Francesca and Fabbietti, Laura and Pinto, Chiara and Singh, Bhawani and Tripathy, Sushanta",
    title = "{A realistic coalescence model for deuteron production}",
    doi = "10.1140/epjc/s10052-023-11972-3",
    journal = "Eur. Phys. J. C",
    volume = "83",
    OPTnumber = "9",
    pages = "804",
    year = "2023"
}

@article{ALICE:2025byl,
    author = {{ALICE Collaboration}},
    title = "{Observation of deuteron and antideuteron formation from resonance-decay nucleons}",
    reportNumber = "CERN-EP-2025-081",
    doi = "10.1038/s41586-025-09775-5",
    journal = "Nature",
    volume = "648",
    OPTnumber = "8093",
    pages = "306--311",
    year = "2025"
}

@article{Andronic:2010qu,
    author = "Andronic, A. and Braun-Munzinger, P. and Stachel, J. and Stocker, H.",
    title = "{Production of light nuclei, hypernuclei and their antiparticles in relativistic nuclear collisions}",
    doi = "10.1016/j.physletb.2011.01.053",
    journal = "Phys. Lett. B",
    volume = "697",
    OPTnumber = "3",
    pages = "203--207",
    year = "2011"
}

@article{Cleymans:2011pe,
    author = "Cleymans, J. and Kabana, S. and Kraus, I. and Oeschler, H. and Redlich, K. and Sharma, N.",
    title = "{Antimatter production in proton-proton and heavy-ion collisions at ultrarelativistic energies}",
    doi = "10.1103/PhysRevC.84.054916",
    journal = "Phys. Rev. C",
    volume = "84",
    OPTnumber = "5",
    pages = "054916",
    year = "2011"
}

@article{Vovchenko:2019kes,
    author = {Vovchenko, Volodymyr and D{\"o}nigus, Benjamin and Stoecker, Horst},
    title = "{Canonical statistical model analysis of $p$-$p$, $p$-Pb, and Pb-Pb collisions at energies available at the CERN Large Hadron Collider}",
    doi = "10.1103/PhysRevC.100.054906",
    journal = "Phys. Rev. C",
    volume = "100",
    OPTnumber = "5",
    pages = "054906",
    year = "2019"
}

@article{Sjostrand:2014zea,
    author = {Sj{\"o}strand, Torbj{\"o}rn and Ask, Stefan and Christiansen, Jesper R. and Corke, Richard and Desai, Nishita and Ilten, Philip and Mrenna, Stephen and Prestel, Stefan and Rasmussen, Christine O. and Skands, Peter Z.},
    title = "{An introduction to PYTHIA 8.2}",
    reportNumber = "LU-TP-14-36, MCNET-14-22, CERN-PH-TH-2014-190, FERMILAB-PUB-14-316-CD, DESY-14-178, SLAC-PUB-16122",
    doi = "10.1016/j.cpc.2015.01.024",
    journal = "Comput. Phys. Commun.",
    volume = "191",
    pages = "159--177",
    year = "2015"
}

@article{Pierog:2013ria,
    author = "Pierog, T. and Karpenko, Iu. and Katzy, J. M. and Yatsenko, E. and Werner, K.",
    title = "{EPOS LHC: Test of collective hadronization with data measured at the CERN Large Hadron Collider}",
    reportNumber = "DESY-13-125",
    doi = "10.1103/PhysRevC.92.034906",
    journal = "Phys. Rev. C",
    volume = "92",
    OPTnumber = "3",
    pages = "034906",
    year = "2015"
}

@article{ALICE:2015ial,
    author = {{ALICE Collaboration}},
    title = "{Measurement of pion, kaon and proton production in proton{\textendash}proton collisions at $\sqrt{s} = 7$ TeV}",
    reportNumber = "CERN-PH-EP-2015-068",
    doi = "10.1140/epjc/s10052-015-3422-9",
    journal = "Eur. Phys. J. C",
    volume = "75",
    OPTnumber = "5",
    pages = "226",
    year = "2015"
}

@article{ALICE:2020nkc,
    author = {{ALICE Collaboration}},
    title = "{Multiplicity dependence of $\pi$, K, and p production in pp collisions at $\sqrt{s} = 13$ TeV}",
    reportNumber = "CERN-EP-2020-024",
    doi = "10.1140/epjc/s10052-020-8125-1",
    journal = "Eur. Phys. J. C",
    volume = "80",
    OPTnumber = "8",
    pages = "693",
    year = "2020"
}

@article{Bhattacharyya:2015hya,
    author = "Bhattacharyya, Trambak and Cleymans, Jean and Khuntia, Arvind and Pareek, Pooja and Sahoo, Raghunath",
    title = "{Radial flow in non-extensive thermodynamics and study of particle spectra at LHC in the limit of small $(q-1)$}",
    doi = "10.1140/epja/i2016-16030-5",
    journal = "Eur. Phys. J. A",
    volume = "52",
    OPTnumber = "2",
    pages = "30",
    year = "2016"
}

@article{Cleymans:2011in,
    author = "Cleymans, J. and Worku, D.",
    title = "{The Tsallis distribution in proton-proton collisions at $\sqrt{s}$ = 0.9 TeV at the LHC}",
    doi = "10.1088/0954-3899/39/2/025006",
    journal = "J. Phys. G Nucl. Partic.",
    volume = "39",
    OPTnumber = "2",
    pages = "025006",
    year = "2012"
}

@article{CMS:2010wcx,
    author = {{CMS Collaboration}},
    title = "{Transverse momentum and pseudorapidity distributions of charged hadrons in pp collisions at $\sqrt{s} = 0.9$ and 2.36 TeV}",
    reportNumber = "CMS-QCD-09-010, CERN-PH-EP-2010-003",
    doi = "10.1007/JHEP02(2010)041",
    journal = "J. High Energy Phys.",
    volume = "2010",
    pages = "041",
    year = "2010"
}

@article{ALICE:2010syw,
    author = "{ALICE Collaboration}",
    title = {{Transverse momentum spectra of charged particles in proton-proton collisions at $\sqrt{s}=900$ GeV with ALICE at the LHC}},
    doi = "10.1016/j.physletb.2010.08.026",
    journal = "Phys. Lett. B",
    volume = "693",
    OPTnumber = "2",
    pages = "53--68",
    year = "2010"
}

@book{Tsallis:2009zex,
    author = "Tsallis, Constantino",
    title = "Introduction to Nonextensive Statistical Mechanics: Approaching a Complex World",
    doi = "10.1007/978-0-387-85359-8",
    isbn = "978-0-387-85358-1, 978-0-387-85359-8",
    publisher = "Springer",
    address = "New York",
    year = "2009"
}

@article{Tsallis:1987eu,
	title = {Possible generalization of {Boltzmann}-{Gibbs} statistics},
	volume = {52},
	issn = {1572-9613},
    reportNumber = "CBPF-NF-062-87",
	doi = {10.1007/BF01016429},
	OPTnumber = {1--2},
    journal = "J. Stat. Phys.",
	author = {Tsallis, Constantino},
	month = jul,
	year = {1988},
	pages = {479--487}
}

@article{Weber2019,
  author  = {Weber, Juliane and Reyers, Mark and Beck, Christian
             and Timme, Marc and Pinto, Joaquim G.
             and Witthaut, Dirk and Sch{\"a}fer, Benjamin},
  title   = "{Wind Power Persistence Characterized by Superstatistics}",
  journal = {Sci. Rep.},
  volume  = {9},
  OPTnumber = {1},
  pages   = {19971},
  year    = {2019},
  doi     = {10.1038/s41598-019-56286-1}
}

@article{Antonopoulos2014,
  author  = {Antonopoulos, Chris G. and Michas, George
             and Vallianatos, Filippos and Bountis, Tassos},
  title   = {Evidence of $q$-exponential statistics in {Greek} seismicity},
  journal = {Physica A},
  volume  = {409},
  pages   = {71--77},
  year    = {2014},
  doi     = {10.1016/j.physa.2014.04.042}
}

@article{Briggs2007,
  author  = {Briggs, Keith and Beck, Christian},
  title   = {Modelling train delays with $q$-exponential functions},
  journal = {Physica A},
  volume  = {378},
  OPTnumber = {2},
  pages   = {498--504},
  year    = {2007},
  doi     = {10.1016/j.physa.2006.11.084}
}

@article{RosalesHerrera:2026pre,
  title = {Complexity drives the symmetry breaking of temperature fluctuations},
  author = {Rosales Herrera, D. and Gallegos Ruiz, M. Y. and Fern{\'a}ndez T{\'e}llez, A. and Alvarado Garc{\'{\i}}a, J. R. and Ram{\'{\i}}rez, J. E.},
  journal = {Phys. Rev. E},
  volume = {114},
  OPTnumber = {4},
  pages = {044117},
  year = {2026},
  month = {Oct},
  publisher = {American Physical Society},
  doi = {10.1103/hmzl-jh2h}
}

@book{Gradshteyn2007,
  author    = {Gradshteyn, I. S. and Ryzhik, I. M.},
  editor    = {Jeffrey, Alan and Zwillinger, Daniel},
  title     = {Table of Integrals, Series, and Products},
  edition   = {7},
  publisher = {Academic Press},
  address   = {Amsterdam},
  year      = {2007},
  isbn      = {978-0-12-373637-6}
}

@article{RosalesHerrera:2025zqt,
    author = "Rosales Herrera, D. and Alvarado Garc{\'\i}a, J. R. and Fern{\'a}ndez T{\'e}llez, A. and Cuautle, E. and Ram{\'\i}rez, J. E.",
    title = "{Scaling laws for softened hadron production at LHC energies}",
    doi = "10.1140/epjc/s10052-025-14469-3",
    journal = "Eur. Phys. J. C",
    volume = "85",
    OPTnumber = "7",
    pages = "760",
    year = "2025"
}

\end{document}